\documentclass[prl,twoside,preprintnumbers,superscriptaddress,twocolumn,nofootinbib,nobibnotes]{revtex4}
\usepackage{hyperref}
\usepackage{amsmath,graphicx,slashed,color}
\usepackage{graphicx,epsfig,amssymb,bm}
\usepackage[utf8]{inputenc}
\def \beq{\begin{equation}}
\def \eeq{\end{equation}}
\def \beqa{\begin{eqnarray}}
\def \eeqa{\end{eqnarray}}

\usepackage{ragged2e}

\usepackage{booktabs}

\usepackage{tabularx}
\newcolumntype{C}[1]{>{\centering\arraybackslash}p{#1}}

\usepackage{graphicx}
\usepackage{url}
\usepackage{bm}
\usepackage{hyperref}
\usepackage{amsmath}
\usepackage{amssymb}
\usepackage{listings}
\usepackage{color}

\usepackage{tabularx}
\newcolumntype{C}[1]{>{\centering\arraybackslash}p{#1}}

\def\bea{\begin{eqnarray}}
\def\eea{\end{eqnarray}}

\def\gsim{\mathrel{\rlap{\lower4pt\hbox{\hskip1pt$\sim$}}
    \raise1pt\hbox{$>$}}}         %greater than or approx. symbol
\def\lsim{\mathrel{\rlap{\lower4pt\hbox{\hskip1pt$\sim$}}
    \raise1pt\hbox{$<$}}}         %less than or approx. symbol

\newcommand{\be}{\begin{equation}}
\newcommand{\ee}{\end{equation}}
\newcommand{\bi}{\begin{itemize}}
\newcommand{\ei}{\end{itemize}}
\newcommand{\ben}{\begin{enumerate}}
\newcommand{\een}{\end{enumerate}}

\usepackage{xcolor}
\usepackage[utf8]{inputenc}
\graphicspath{{./figs/}}

\usepackage{amssymb}
\usepackage{dcolumn}	% Align table columns on decimal point
\usepackage{bm}			% bold math
\usepackage{bbm} 		% blacboard math for \mathbbm{1}
\usepackage{multirow}
\usepackage{slashed}
\usepackage{blkarray}
\usepackage{booktabs}
\usepackage{makecell}
\usepackage{tikz}
\usepackage[compat=1.1.0]{tikz-feynman}
\usepackage{placeins}
\usepackage{listings}
\usepackage{amsmath}
\usepackage{xcolor}
\usepackage{pythonhighlight}
\usepackage{subcaption} 

\def\gsim{\mathrel{\rlap{\lower4pt\hbox{\hskip1pt$\sim$}}
    \raise1pt\hbox{$>$}}}         %greater than or approx. symbol
\def\lsim{\mathrel{\rlap{\lower4pt\hbox{\hskip1pt$\sim$}}
    \raise1pt\hbox{$<$}}}         %less than or approx. symbol

\definecolor{codegreen}{rgb}{0,0.6,0}
\definecolor{codegray}{rgb}{0.5,0.5,0.5}
\definecolor{codepurple}{rgb}{0.58,0,0.82}
\definecolor{backcolour}{rgb}{0.95,0.95,0.92}

\lstdefinestyle{mystyle}{
    backgroundcolor=\color{backcolour},   
    commentstyle=\color{codegreen},
    keywordstyle=\color{magenta},
    numberstyle=\tiny\color{codegray},
    stringstyle=\color{codepurple},
    basicstyle=\ttfamily\footnotesize,
    breakatwhitespace=false,         
    breaklines=true,                 
    captionpos=b,                    
    keepspaces=true,                 
    numbers=left,                    
    numbersep=5pt,                  
    showspaces=false,                
    showstringspaces=false,
    showtabs=false,                  
    tabsize=2
}
\definecolor{Red}{rgb}{1.,0.,0.}

\usepackage[normalem]{ulem}
\definecolor{ao}{rgb}{0.0, 0.5, 0.0}
\definecolor{mt}{RGB}{108, 210, 92}
\newcommand{\bc}{\begin{center}}
\newcommand{\ec}{\end{center}}
\newcommand{\ba}{\begin{array}}
\newcommand{\ea}{\end{array}}

\newcommand{\GeV}{\,\mathrm{GeV}}

\usepackage{tabularx}

\usepackage[normalem]{ulem}
\definecolor{ao}{rgb}{0.0, 0.5, 0.0}
\definecolor{mt}{RGB}{108, 210, 92}

\begin{document}
\title{Determining the Threshold Excess in Top-Quark Pair Production through a Global PDF Analysis }

\author{Richard~D.~Ball}
\author{Jaco~ter~Hoeve}
\affiliation{The Higgs Centre for Theoretical Physics, University of Edinburgh, 
Mayfield Road, Edinburgh EH9 3JZ, United Kingdom}

\date{\today}

\begin{abstract}
We present an empirical determination of the threshold excess $\Delta$ in top-quark pair production, taking full account of correlated uncertainties with PDFs, the 
top-quark mass $m_t$ and the strong
coupling $\alpha_s(m_Z)$. We extract $\Delta$ from a variety of differential $t\bar{t}$ cross-section
measurements by ATLAS and CMS at 8 and 13 TeV, using theoretical predictions of the $t\bar{t}$ continuum up
to NNLO QCD accuracy, including an estimate of missing higher order uncertainties and NLO EW
corrections. At 8 TeV we see no evidence of a significant threshold excess, but 
at 13 TeV obtain results above the NNLO QCD $t\bar{t}$ background at just over two $\sigma$ when we combine all the available measurements. We also find evidence of a strong correlation between $\Delta$ and $m_t$, which must be taken into account if the uncertainty on any threshold excess is to be estimated reliably. 
\end{abstract}

\maketitle

\noindent
{\bf Introduction.}
The top quark occupies a unique place in the Standard Model (SM) as the heaviest
known elementary particle. Its pole mass $m_t$ provides key input to many LHC
processes, and together with the strong coupling constant $\alpha_s$ they
determine the running of the Higgs self-coupling responsible for determining
whether our Universe resides in an absolute or meta-stable state. The ATLAS and
CMS collaborations at the LHC have measured $t\bar{t}$ production extensively in
a large series of differential cross-section measurements at 8 and 13 TeV
~\cite{ATLAS:2022jbj,ATLAS:2020ccu,ATLAS:2019hxz,ATLAS:2016pal,ATLAS:2015lsn,CMS:2024ybg,CMS:2021vhb,CMS:2018adi},
providing a rich data set from which to determine the top-quark mass. \\
\indent An intriguing feature of top-quark pair production near threshold is the
possible formation of a colour-singlet $t\bar{t}$ quasi-bound state, known as
toponium, just below the $t\bar{t}$ production threshold. This state manifests
as a localised excess in the $t\bar{t}$ invariant mass distribution
$m_{t\bar{t}}$ just below $2m_t$. Here,  NLO NRQCD calculations predict a contribution to
the total cross-section of $4.15^{+1.43}_{-1.47}$ pb with respect to the NLO QCD $t\bar{t}$ continuum for an invariant mass cut at $m_{t\bar{t}}=350$ GeV and unstable tops \cite{Garzelli:2026ctb}, while Ref.~\cite{Nason:2026oka} instead predicts an excess of $3.87(2)$ pb for the same $m_{t\bar{t}}$ cut, which is conjectured to further drop to around two picobarns in NNLO QCD. The ATLAS and CMS collaborations have recently provided
the first evidence of an excess, reporting toponium cross-sections
of $9.3^{+1.4}_{-1.3}$ pb~\cite{ATLAS:2026dbe} and $8.8^{+1.2}_{-1.4}$
pb~\cite{CMS:2025kzt} at $\sqrt{s}=13$ TeV, respectively, in tension with
the NRQCD predictions. This may point to incomplete
theoretical control of the calculations~\cite{Nason:2025hix,Nason:2026oka, Garzelli:2026ctb},
since this requires careful subtraction of perturbative QCD contributions that
overlap with the all-orders Coulomb resummation.\\
\indent In Ref.~\cite{Ball:2026qno}, we performed a joint determination of the pole mass 
$m_t$ and $\alpha_s(m_Z)$ in the context of a global PDF fit, while accounting for the impact
of a possible threshold excess $\Delta$ on the value of $m_t$ with respect to fixed order NNLO QCD complemented with EW corrections. In this work, we instead treat $\Delta$ as a free parameter, to be extracted simultaneously with $m_t$ and $\alpha_s(m_Z)$, alongside the PDFs, thereby taking full account of correlations between all these quantities. We thus take a purely data-driven approach and do not attempt to associate the excess with any particular theoretical model.

\vspace{0.1cm}
\noindent
{\bf Methodology.}
Our approach is based on the Theory Covariance Method (TCM) as developed in
Refs.~\cite{Ball:2018lag,Ball:2018twp,NNPDF:2019ubu,Ball:2021icz} and applied to
the joint extraction of $m_t$ and $\alpha_s(m_Z)$ in Ref.~\cite{Ball:2026qno}.
We refer to Ref.~\cite{Ball:2026qno} for an explanation of the method, and here indicate only how we extended the formalism to also determine the threshold excess. 

The TCM can be used to determine the posterior distribution of any nuisance
parameter $\lambda_a$ given a set of $n_{\rm dat}$ experimental measurements $D$ and corresponding
theoretical predictions $T$, with Gaussian uncertainties summarised in a combined covariance matrix $C$. The nuisance parameters
$\lambda_a$ centered around zero induce a 
shift $\beta_a$ on the theory
predictions $T \rightarrow T + \lambda_a \beta_a$. Assuming a unit Gaussian prior for the nuisance parameters, the posterior
$P(\lambda | TD)$ is again Gaussian with mean and covariance given by
\begin{align}
    \nonumber\bar{\lambda}_a &= \beta_a^T(C+S)^{-1}(D-T)\\
    Z_{ab} &= \delta_{ab} - \beta_a^T(C+S)^{-1}\beta_b \, ,
    \label{eq:lambda_post}
\end{align}
where $S$ is constructed from the shifts $\beta_a$ by $S=\beta_a \beta_a^T$.
Applying Eq.~\eqref{eq:lambda_post} repeatedly to a set of data replicas $D^{(r)}$,
\begin{equation}
    D^{(r)} \sim \mathcal{N}(D, C+S) \, ,
\end{equation}
with corresponding theoretical predictions $T^{(r)}$ leads to a replica ensemble
$\lambda_a^{(r)}$ with mean and covariance given by
\begin{align}
    \nonumber \langle \bar{\lambda_a} \rangle &= \beta^T(C+S)^{-1}(D-T^{(0)}) \, \\
    \nonumber \bar{Z}_{ab} &\equiv \delta_{ab} - \beta_a^T(C+S)^{-1}\beta_b \\
    &\hspace{2cm} +  \beta_a^T(C+S)^{-1}X(C+S)^{-1}\beta_b \, ,
    \label{eq:lambda_post_reps}
\end{align}
where the angular brackets denote an average over replicas, while $X$ denotes
the covariance of the theory predictions $T^{(r)}$ taken over replicas.

This formalism can also be used to constrain  theoretical parameters
$\tilde{T}$ that have not been fitted to $D$ while having theoretical uncertainties
parameterised by $\tilde{\beta}_a$ that are fully correlated to those of $T$,
i.e.
\begin{equation}
\tilde{T} \rightarrow \tilde{T} + \lambda_a \tilde{\beta}_a \, .
\end{equation}
It then follows from Eq.~\eqref{eq:lambda_post_reps} that the prior central value
$\tilde{T}$ receives a shift
\begin{equation}
    \delta \tilde{T} = \hat{S}(C+S)^{-1}(D-T^{(0)}) \, ,
    \label{eq:cv_shift}
\end{equation}
where we have defined $\hat{S} \equiv \tilde{\beta}_a\beta_a^T$ as the
cross-covariance between $T$ and $\tilde{T}$. The corresponding covariance is
given by
\begin{align}
    \nonumber{\rm Cov}[\tilde{T}] &\equiv \tilde{S} - \hat{S}(C+S)^{-1}\hat{S}^T \\
    &\hspace{2cm} + \hat{S}(C+S)^{-1}X(C+S)^{-1}\hat{S}^T \, ,
    \label{eq:cov}
\end{align}
where $\tilde{S} \equiv \tilde{\beta}_a\tilde{\beta}_a^T$ denotes the prior
theoretical covariance.

The application to determining the shift on the prior central values $\tilde{T}
= (m_t^{(0)}, \alpha_s^{(0)}, \Delta^{(0)})$ is straightforward. We define the prior central
values $m_t^0 = 172.5\,$GeV, $\alpha_s^0 = 0.120$, 
$\Delta^{(0)} = 6.43\,$pb ($2.19\,$pb) at 13 (8) TeV as per Ref.~\cite{Fuks:2021xje}, with
prior theoretical uncertainties $\delta m_t = 2.5\,$GeV, $\delta\alpha_s =
0.002$ and $\delta\Delta^{(0)} = 2\Delta^{(0)}$. We deliberately choose a 200\%
relative uncertainty on the threshold excess in order not to bias it towards the
prior central value $\Delta^{(0)}$. With this choice of prior, we have verified that the posterior lies well inside the prior range. To determine the prior uncertainties, theoretical predictions are computed at the six varied parameter points
$\{m_t^\pm, \alpha_s^\pm, \pm 2\Delta^{(0)}\}$, giving a 6-component vector $\beta_a$,
\begin{equation}
    \beta_a = \frac{1}{\sqrt{2}}\left(\delta T_{m_t}^+, \delta T_{m_t}^-, \delta T_{\alpha_s}^+, \delta T_{\alpha_s}^-, 2\Delta^{(0)}, -2\Delta^{(0)}\right) \, ,
\end{equation}
where $\delta T^{\pm} = T^\pm-T^{(0)}$ defines the shift with respect to the central theory predictions $T^{(0)}$. The matrices $\tilde{S}$ and $\hat{S}$ now follow
directly,
\begin{equation}
    \tilde{S} = \frac{1}{2}\begin{pmatrix}
\delta m_t^2 & 0 & 0\\
0 & \delta\alpha_s^2 & 0 \\
0 & 0 & (\delta\Delta^{(0)})^2
\end{pmatrix} \, ,
\label{eq:S_tilde}
\end{equation}
\begin{equation}
    \hat{S}_i = \frac{1}{2}\begin{pmatrix}
(\delta T^+_{m_t,i}-\delta T^-_{m_t,i})\delta m_t \\
(\delta T^+_{\alpha_s,i}-\delta T^-_{\alpha_s,i}) \delta \alpha_s \\
4(\Delta^{(0)})^2
\end{pmatrix} \, .
\label{eq:S_hat}
\end{equation}
Substituting Eqs.~\eqref{eq:S_tilde} and \eqref{eq:S_hat} into Eqs.~\eqref{eq:cv_shift} and \eqref{eq:cov} then gives a posterior estimate in the joint $(m_t, \alpha_s, \Delta)$ space, where $\Delta$ should be understood as an excess with respect to the $t\bar{t}$ fixed order NNLO QCD background, complemented with EW corrections at order  $\mathcal{O}(\alpha_s\alpha)$ and $\mathcal{O}(\alpha_s^2\alpha)$. 

\vspace{0.1cm}
\noindent
{\bf Settings.}
The experimental data set and theory settings are the same as those adopted
in Ref.~\cite{Ball:2026qno}. In particular, we take the NNPDF4.0 data set as our baseline \cite{NNPDF:2021njg} and complement it with new $t\bar{t}$ measurements that have become available since its release, which we show in Table \ref{tab:datasets}. We consider only $t\bar{t}$ measurements differential in the invariant mass $m_{t\bar{t}}$ of the top-quark pair as this is the most sensitive observable around threshold. All $t\bar{t}$ measurements are computed at NNLO QCD accuracy using {\sc MATRIX} interfaced to {\sc PineAPPL}~\cite{Devoto:2025cuf,Grazzini:2017mhc, Carrazza:2020gss} complemented with EW corrections at order $\mathcal{O}(\alpha_s\alpha)$ and $\mathcal{O}(\alpha_s^2\alpha)$ obtained with {\sc MadGraph5\_aMC\@NLO} \cite{Alwall:2014hca}. We neglect bin migrations arising from the finite top-quark width, since these predominantly shift events within the threshold bin, which is much wider than the top-quark width \cite{Garzelli:2026ctb}. We adopt {\sc EKO} for PDF evolution while including QED and approximate N$^3$LO QCD corrections corrections following Ref.~\cite{NNPDF:2024djq} and Refs.~\cite{NNPDF:2024nan, Ablinger:2024xtt, Falcioni:2024qpd}, respectively. We include missing higher-order uncertainties (MHOUs) throughout following the prescription of Ref.~\cite{NNPDF:2024dpb}. 
\vspace{0.1cm}

\begin{figure}
    \centering
    \includegraphics[width=\linewidth]{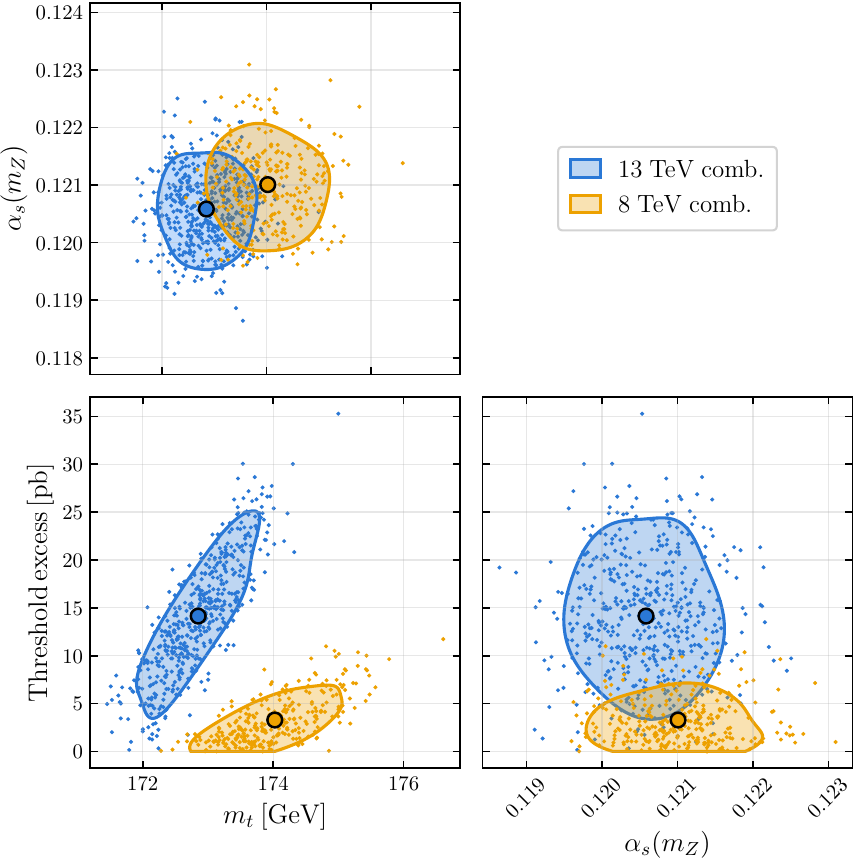}
    \caption{68\% C.L bounds on the threshold excess $\Delta$, the top-quark mass $m_t$ and the strong coupling $\alpha_s(m_Z)$. These are combined fits to all ATLAS and CMS top datasets, at either 
    $\sqrt{s}=8$ TeV or $\sqrt{s}=13$ TeV. The dots show the results for individual PDF replicas, which together determine the 68\% C.L. regions.}
    \label{fig:cornerplot}
\end{figure}

%-------------------------------------------------------------------------------
\begin{table}[t]
  
    \renewcommand{\arraystretch}{1.40}
    \centering
  \begin{tabular}{|c|c|c|c|}
\hline
\bf{data set}
& $n_{\rm dat}$
& $\mathcal{L}_{\rm int}$ $[{\rm fb}^{-1}]$
& \bf{Ref.}
\\
\hline
ATLAS 13~TeV $t\bar{t}$ all hadr. &  9 &36.1  & \cite{ATLAS:2020ccu}\\
\hline
ATLAS 13~TeV $t\bar{t}$ $\ell+j$ & 9 & 36.0 & \cite{ATLAS:2019hxz} \\
\hline
ATLAS 8~TeV $t\bar{t}$ $2\ell$ & 6 & 20.2 & \cite{ATLAS:2016pal} \\
\hline
ATLAS 8~TeV $t\bar{t}$ $\ell+j$  & 7   & 20.3 & \cite{ATLAS:2015lsn} \\
\hline
CMS 13~TeV $t\bar{t}$ $2\ell$  & 7   &138.0  & \cite{CMS:2024ybg}\\
\hline
CMS 13~TeV $t\bar{t}$ $\ell+j$& 15  & 137.0 & \cite{CMS:2021vhb} \\
\hline
\end{tabular}

  \caption{Overview of the $t\bar{t}$ measurements differential in the invariant mass $m_{t\bar{t}}$ considered in the current
  analysis. For each data set, we indicate the
  number of data points, $n_{\rm dat}$, the integrated luminosity,
  $\mathcal{L}_{\rm int}$ and the corresponding reference.}
  \label{tab:datasets}
\end{table}
%------------------------------------------------------------------------------
\vspace{0.1cm}
\noindent
{\bf Results.}
Fig.~\ref{fig:cornerplot} shows the best-fit values of the top-quark mass $m_t$, the strong coupling $\alpha_s(m_Z)$ and the threshold excess $\Delta$ with respect to the $t\bar{t}$ continuum and for each PDF replica entering Eq.~\eqref{eq:cv_shift}. Results are shown both for a threshold excess determination at $\sqrt{s}=8$ TeV and $\sqrt{s}=13$ TeV. Given that the threshold excess when interpreted in terms of a quasi-bound state enhances the cross-section, we discard PDF replicas that lead to $\Delta < 0$ and show the 68\% C.L. bounds on the remaining replicas obtained through a Kernel Density Estimate (KDE). The numerically  marginalised bounds are collected in Table \ref{tab:cl68_bounds}.

Focusing first on the combination at $\sqrt{s}=13$ TeV with $\Delta > 0$, we find a non-zero threshold excess with $68\%$ C.L bounds 
\begin{equation}
\Delta =  14.14^{+6.44}_{-6.31}\,{\rm pb}\, , \quad
m_t = 172.85^{+0.59}_{-0.58}\,\GeV \, ,
\end{equation}
which is compatible with the $t\bar{t}$ continuum within 2.2$\sigma$. The resulting $m_t$ determination is consistent with our previous determination from Ref.~\cite{Ball:2026qno}, although its uncertainty is larger by a factor of $2.3$ as both $m_t$ and $\Delta$ are now allowed to vary. We find a high correlation coefficient between  $\Delta$ and $m_t$ of $\rho=0.84$, a consequence of the fact that the same cross-section can be obtained for a higher top-quark mass when increasing the threshold excess. This kinematic interplay between
the magnitude of the excess and the position of the threshold produces the elongated
posterior contour in the $(\Delta, m_t)$ plane, which indicates that threshold determinations must be performed jointly with $m_t$. We find instead a negligible correlation between $\alpha_s(m_Z)$ and the threshold excess. This is expected since the threshold excess is a
localised kinematic effect near $m_{t\bar{t}} \approx 2m_t$, while $\alpha_s$ is
sensitive to the normalisation and shape of the cross-section across all
kinematic bins and is also constrained by other data in the global fit.

This is to be contrasted with the combination at $\sqrt{s}=8$ TeV, where the threshold excess is reduced significantly with respect to the determination at $\sqrt{s}=13$ TeV, again while imposing positivity, $\Delta >0 $, 
\begin{equation}
\Delta =  3.28^{+2.66}_{-2.32}\,{\rm pb}\, , \quad
m_t = 174.02^{+0.71}_{-0.66}\,\GeV \, .
\label{eq:bound_8_tev_pos}
\end{equation}
The correlation coefficient between $\Delta$ and $m_t$ remains large, however, at $\rho=0.69$. While the scale associated with the threshold excess is fixed around $2m_t$, as one moves to lower $\sqrt{s}$, the gluons responsible for quasi-bound state formations will have higher values of Bj\"orken-$x$. Since the gluon PDF falls steeply with rising $x$, this then decreases the threshold excess at 8 TeV compared to 13 TeV. Due to the reduced excess at 8 TeV, the resulting marginalised 68\% C.L. bounds now depend significantly on whether positivity is imposed or not, in contrast to the $13$ TeV combination. Indeed, the corresponding bound reported in  Eq.~\eqref{eq:bound_8_tev_pos} in the absence of positivity is significantly lower for both $\Delta$ and $m_t$,
\begin{equation}
    \Delta =  0.57 \pm 3.83\,{\rm pb}\, , \quad
m_t = 173.53 \pm 0.85\,\GeV \, ,
\label{eq:bound_8_tev_no_pos}
\end{equation}
which follows directly from the positive correlation between $m_t$ and the replicas with negative excess. 

\begin{figure}
    \centering
    \includegraphics[width=.8\linewidth]{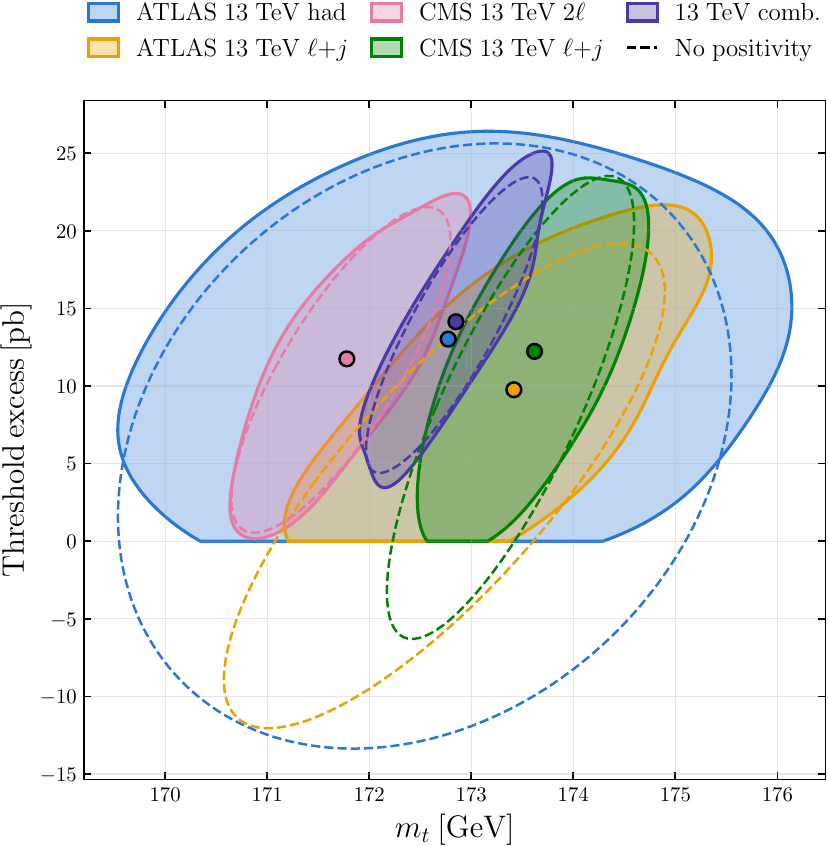}
    \caption{The 68\% C.L. bounds on the top-quark mass $m_t$ and the threshold excess with respect to the $t\bar{t}$ continuum at NNLO QCD at 13 TeV, shown for each data set separately as well as their combination.}
    \label{fig:cl_bounds_13tev}
\end{figure}

\begin{figure}[!h]
    \centering
    \includegraphics[width=\linewidth]{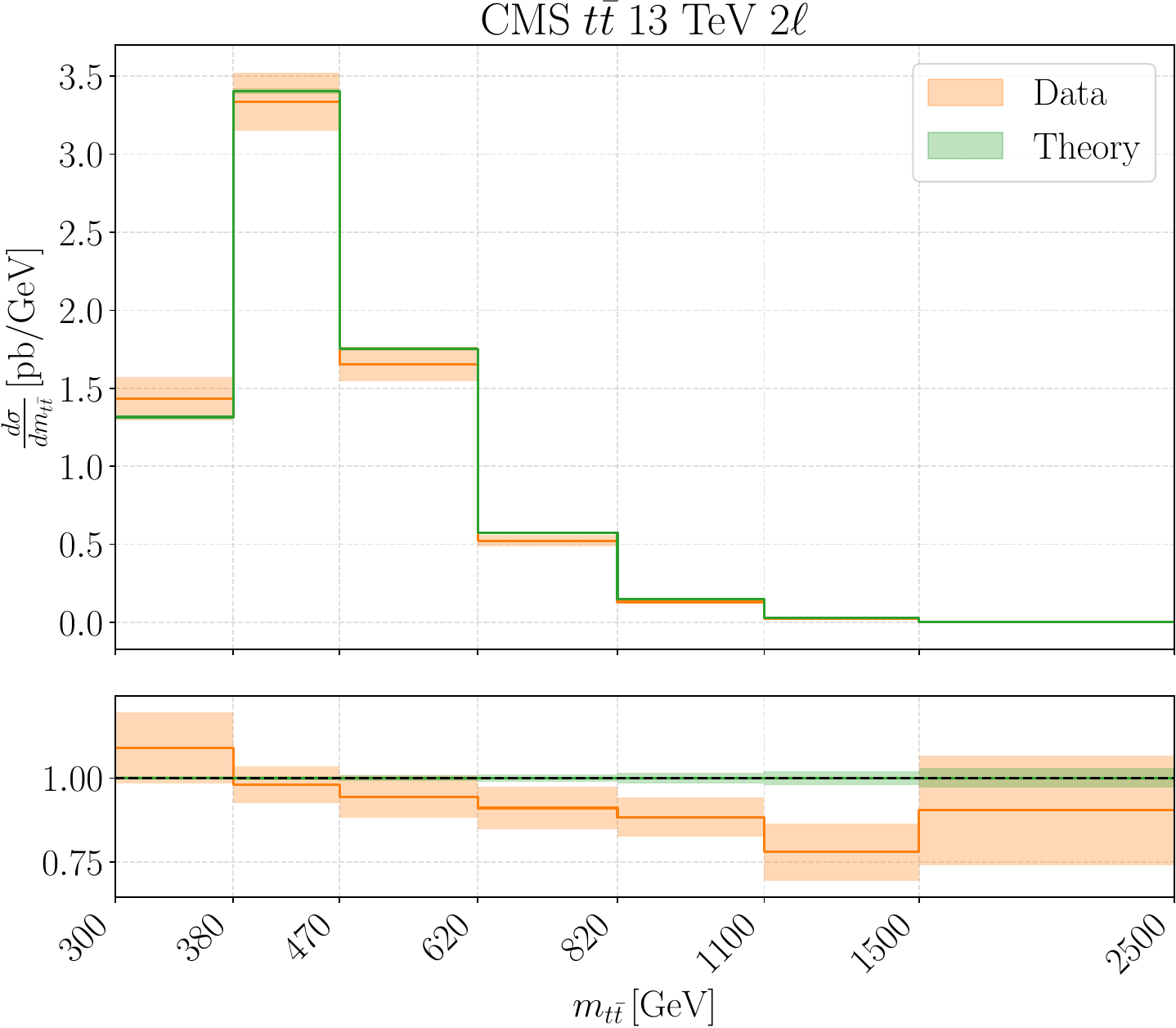}
    \caption{Data theory comparison of the CMS $t\bar{t}$ measurement at 13 TeV in the dilepton channel from Ref.~\cite{CMS:2024ybg}. The theory has been fitted at NNLO QCD with EW corrections while including MHOUs.}
    \label{fig:data_theory}
\end{figure}

Fig.~\ref{fig:cl_bounds_13tev} shows the breakdown of the threshold determination at $\sqrt{s}=13$ TeV in terms of the individual $t\bar{t}$ datasets that enter the combination. In addition to the 68\% C.L. obtained from a KDE estimate of PDF replicas yielding a positive excess, we now also present the resulting 68\% C.L. ellipses (dotted) obtained when PDF replicas corresponding to a negative excess are also included. Note that the two-dimensional 68\% C.L. intervals are wider along each axis than the corresponding one-dimensional one by a factor $\sqrt{r}\approx 1.51$ with $r = \Delta \chi^2_{2,{\rm dof}}$. Several observations emerge from Fig.~\ref{fig:cl_bounds_13tev}. First, while the $m_t$
values do not fully overlap, the threshold excess is
consistent within uncertainties between all data sets, with the purely hadronic channel leading to the largest uncertainties due to its relatively large experimental systematics. We also note that the $\ell + j$ channel gives consistent results between ATLAS and CMS in $m_t$, while there is a slight tension between the $\ell+j$ and $2\ell$ channels in CMS. Another notable feature is that the central value of the combination is consistently higher than those of the individual determinations. Although counterintuitive, this arises because the separate data sets are correlated through a common nuisance parameter and combining correlated measurements can produce this behaviour, as pointed out in the context of the BLUE method \cite{Lyons:1988rp}. Fig.~\ref{fig:cl_bounds_13tev} further shows a $2.2\sigma$ tension with the background only hypothesis for the combination, while each individual data set predicts an excess compatible with zero within uncertainties, except for the CMS measurement in the $2\ell$ channel. We highlight the data theory comparison for this dataset in Fig.~\ref{fig:data_theory}. We find that the PDF fit tends to undershoot the data in the threshold region, while overshooting it well above threshold. This effect is then compensated in the TCM formalism either by an increased threshold excess, or a reduced top-quark mass. Indeed, in Fig.~\ref{fig:cl_bounds_13tev}, we find that this dataset predicts a clear threshold excess away from zero, at the two sigma level (see also Table \ref{tab:cl68_bounds}), and a relatively low value of $m_t$ compared to the other datasets. 

\begin{figure}
    \centering
    \includegraphics[width=.8\linewidth]{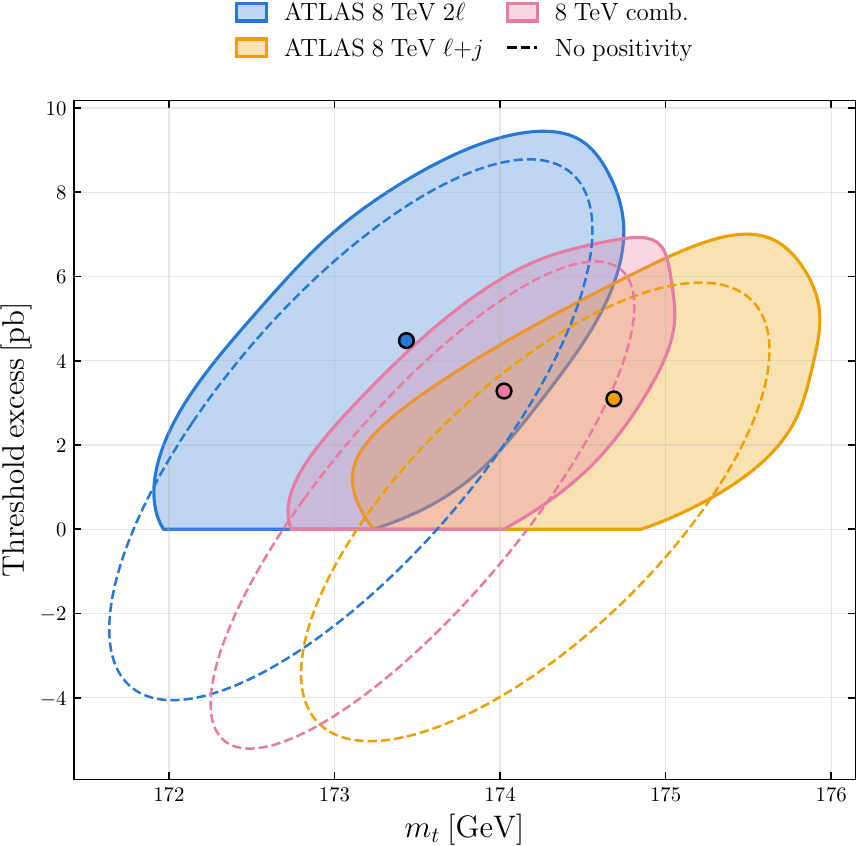}
    \caption{The 68\% C.L. bounds on the top-quark mass $m_t$ and the threshold excess with respect to the $t\bar{t}$ continuum at NNLO QCD at 8 TeV, shown for each data set separately as well as their combination.}
    \label{fig:cl_bounds_8tev}
\end{figure}

Fig.~\ref{fig:cl_bounds_8tev} presents the equivalent breakdown now for the $\sqrt{s}=8$ TeV combination. Although the $2\ell$ and the $\ell + j$ channels from ATLAS give consistent results for $m_t$ and the threshold excess, both are compatible with the $t\bar{t}$ continuum within $1\sigma$. Therefore, no definite statement regarding the formation of a quasi-bound state can be made from the 8 TeV data.

\begin{table}[t]
    \renewcommand{\arraystretch}{1.40}
    \centering
    \begin{tabular}{|c|c|c|c|}
        \hline
        \textbf{Data set} & $\Delta$ [pb] & $\Delta>0$ [pb] & $\chi^2$ \\
        \hline\hline
        % \multicolumn{4}{l}{$13\:{\rm TeV}$} \\
        % \hline
        ATLAS 13 TeV $t\bar{t}$ all hadr. & $6.13 \pm 12.90$ & $13.02^{+8.33}_{-8.82}$ & 1.15 \\
        \hline
        ATLAS 13 TeV $t\bar{t}$ $\ell+j$ & $3.56 \pm 10.33$ & $9.76^{+7.64}_{-7.18}$ & 0.85 \\
        \hline
        CMS 13 TeV $t\bar{t}$ 2$\ell$ & $11.04 \pm 6.94$ & $11.75^{+6.50}_{-6.87}$ & 1.71 \\
        \hline
        CMS 13 TeV $t\bar{t}$ $\ell + j$ & $8.62 \pm 9.87$ & $12.23^{+7.26}_{-7.55}$ & 1.04 \\
        \hline
        13 TeV combination & $13.92 \pm 6.31$ & $14.14^{+6.44}_{-6.31}$ & 1.09 \\
        \hline\hline
        % \multicolumn{4}{l}{$8\:{\rm TeV}$} \\
        ATLAS 8 TeV $t\bar{t}$ 2$\ell$ & $2.36 \pm 4.25$ & $4.48^{+3.04}_{-3.17}$ & 0.08 \\
        \hline
        ATLAS 8 TeV $t\bar{t}$ $\ell + j$ & $0.41 \pm 3.60$ & $3.09^{+2.05}_{-2.12}$ & 1.07 \\
        \hline
        8 TeV combination & $0.57 \pm 3.83$ & $3.28^{+2.66}_{-2.32}$ & 0.83 \\
        \hline
    \end{tabular}
    \caption{Marginalised 68\% CL intervals for the threshold excess, with and
    without the positivity requirement $\Delta>0$ imposed on the replicas,
    together with the fit quality $\chi^2$ per data point, obtained from the
    individual datasets and their combinations at 13 TeV and 8 TeV.}
    \label{tab:cl68_bounds}
\end{table}

In order to try to understand better the relatively small uncertainties in the threshold excess reported by ATLAS and CMS  ($9.3^{+1.4}_{-1.3}$ pb~\cite{ATLAS:2026dbe} and $8.8^{+1.2}_{-1.4}$
pb~\cite{CMS:2025kzt} respectively, both at $\sqrt{s}=13$ TeV and $m_t=172.5$ GeV) we also analysed our results by conditioning on a fixed value of $m_t$, see Table \ref{tab:conditional_topo}. These results are much more similar to those of the experimentalists, though our uncertainties remain rather larger, possibly due to the significant correlation with the PDFs also included in our analysis. 

%In general, we find an excess $\Delta$ at 13 TeV with respect to NNLO QCD that is significantly higher than the two picobarns as predicted in Ref.~\cite{Nason:2026oka}, which suggests that our data-driven approach captures effects beyond a purely perturbative QCD description. 

\begin{table}[t!]
    \renewcommand{\arraystretch}{1.40}
    \centering
    \begin{tabular}{|c|c|c|}
        \hline
        $m_t\:[{\rm GeV}]$ & 13 TeV combination & 8 TeV combination \\
        \hline
        $172.0$ & $8.26^{+4.14}_{-3.76}$ & - \\
        \hline
        $172.5$ & $11.14^{+4.08}_{-4.41}$ & $0.88^{+0.65}_{-0.74}$ \\
        \hline
        $173.0$ & $15.08^{+4.43}_{-4.33}$ & $1.62^{+0.95}_{-0.97}$ \\
        \hline
        $173.5$ & $19.05^{+4.39}_{-4.07}$ & $2.08^{+1.49}_{-1.43}$ \\
        \hline
        $174.0$ & $22.69^{+3.60}_{-4.12}$ & $2.78^{+1.66}_{-1.84}$ \\
        \hline
    \end{tabular}
    \caption{Conditional 68\% CL bounds on the threshold excess $\Delta > 0 $ for fixed values of $m_t$.}
    \label{tab:conditional_topo}
\end{table}

\vspace{0.1cm}\noindent
{\bf Summary and outlook.}
We have presented a determination of the threshold excess in $t\bar{t}$ production, extracted simultaneously with the top-quark mass and strong coupling through a global PDF fit. The excess is determined with respect to the fixed order $t\bar{t}$ NNLO QCD background complemented with EW corrections and missing higher-order uncertainties. Our combined determination at 13 TeV predicts a non-zero excess in tension with the background by $2.2\sigma$, while no tension is observed in our determination at 8 TeV. However, no significant tension in found in our analysis for any of the individual 13 TeV datasets: it is only in their combination that we see significant excess.

We also find a strong positive correlation between $\Delta$ and $m_t$. This correlation arises because both parameters influence the $t\bar{t}$ cross-section in the threshold region of $m_{t\bar{t}}$, and means that any extraction of $m_t$ that ignores the threshold excess will absorb part of the excess into a biased value of $m_t$. Conversely, experimental determinations of the threshold excess must account for the top-mass sensitivity of the threshold region. A fully consistent treatment, such as the joint extraction presented here, is therefore essential for precision determinations of either quantity. 

We also find a significant dependence of the threshold excess on centre-of mass energy. It might be useful in the future for theoretical predictions of toponium cross-sections to be performed for a range of centre-of-mass energies, in order to test the steep rise in 
$\Delta$ we find when going from 8 to 13 TeV.

Our analysis could be extended by adding spin correlation observables to the combined fit in order to break the strong correlation observed between the top-mass and the threshold excess. Furthermore, it would be interesting to repeat the analysis under HL-LHC conditions to examine how the reported excess might be affected by improved experimental resolution in the threshold region.

\noindent
\vspace{0.1cm}
\textit{Acknowledgements.}
We thank the members of the NNPDF collaboration for insightful discussions during the course of this work and are particularly grateful to the members of the Top WG who inspired this study. 
The work of R.D.B. and J. t. H. is supported by the
UK STFC consolidated grant ST/X000494/1.

\bibliography{toponium-mtop}

\end{document}